\documentstyle[12pt]{article}
\begin{document}
\title{Relativistic Hartree-Bogoliubov description\\
of sizes and shapes of $A=20$ isobars}
\author{G.A. Lalazissis$^{1,2}$, D. Vretenar$^{1,3}$, and P. Ring$^{1}$
\vspace{0.5 cm}\\
$^{1}$ Physik-Department der Technischen Universit\"at M\"unchen,\\
D-85748 Garching, Germany\\
$^{3}$ Physics Department, Aristotle University of Thessaloniki,\\
Thessaloniki GR-54006, Greece\\
$^{3}$ Physics Department, Faculty of Science, 
\\University of Zagreb, 10 000 Zagreb, Croatia}
\maketitle
\bigskip
\bigskip
\begin{abstract}
Ground-state properties of $A = 20$ nuclei
($^{20}$N, $^{20}$O, $^{20}$F, $^{20}$Ne, $^{20}$Na, $^{20}$Mg)
are described in the framework
of Relativistic Hartree-Bogoliubov (RHB) theory. The model uses
the NL3 effective interaction in the mean-field Lagrangian,
and describes pairing correlations
by the pairing part of the finite range Gogny interaction D1S.
Binding energies, quadrupole deformations, nuclear matter radii,
and differences in radii of proton and neutron distributions are
compared with recent experimental data.
\end{abstract}
\bigskip \bigskip

\vspace{1 cm} {PACS:} {21.60.Jz, 21.10.Dr, 21.10.Gv, 
27.30.+t}\newline
\vspace{1 cm}\newline
\newpage
\baselineskip = 24pt

%
\section{Introduction}
Sizes, shapes and binding energies are fundamental characteristics 
of nuclei and reflect the basic properties of effective nuclear
forces. The description of the generalized moments of the nuclear
density distributions provides an important test for nuclear
structure models. In particular, a description of ground-state
properties of an isobaric sequence of nuclei tests the isovector
channel of the effective nuclear force. The correct parameterization
of this channel is essential for the description of structure
phenomena in exotic nuclei far from $\beta$-stability.
In the present study we analyze the sequence of $A = 20$ nuclei:
$^{20}$N, $^{20}$O, $^{20}$F, $^{20}$Ne, $^{20}$Na, $^{20}$Mg.
Experimental data on nuclear matter radii derived from the 
measured interaction cross section of these nuclei have been
reported in Ref.~\cite{Chul.96}. Evidence has been found for the
existence of a proton skin for $^{20}$Mg and of a neutron skin
for $^{20}$N. The largest difference in radii ($\approx 0.2$ fm)
has been reported for the mirror nuclei $^{20}$O and $^{20}$Mg.
This last result has prompted several theoretical investigations,
but the extremely small value for the matter radius of
$^{20}$O has not been reproduced in calculations which 
include the microscopic three-cluster model~\cite{Des.98},
the Hartree-Fock model with Skyrme 
interactions~\cite{BH.96,KTS.97,SFB.99,SLR.99},
a single particle potential model~\cite{SFB.99}, 
and the shell-model~\cite{SLR.99}. The experimental data on 
interaction cross sections of $A = 20$ isobars were also 
reviewed in Ref.~\cite{KKF.99} by using a variety of phenomenological
and semi-microscopic models.

In the present analysis of the $A = 20$ isobaric sequence
we employ the Relativistic Hartree-Bogoliubov (RHB) theory.
Based on the relativistic mean-field theory and on the 
Hartree-Fock-Bogoliubov framework, the RHB theory provides a 
unified description of mean-field and pairing correlations.
It has been successfully applied in the description of 
nuclear structure phenomena, not only in nuclei along the valley
of $\beta$-stability, but also of exotic nuclei close to the 
particle drip lines. In particular, applications relevant to
the present study include: the halo phenomenon in light nuclei~\cite{PVL.97},
properties of light nuclei near the neutron-drip~\cite{LVP.98},
reduction of the spin-orbit potential in nuclei with extreme
isospin values~\cite{LVR.97},
the deformation and shape coexistence phenomena that 
result from the suppression
of the spherical N=28 shell gap in neutron-rich nuclei~\cite{Lal.99},
properties of proton-rich nuclei and the phenomenon of
ground-state proton radioactivity~\cite{VLR.99,LVR.99}.

The relativistic mean field theory is based on simple concepts~\cite{Rin.96}:
nucleons are described as point particles, the theory is fully Lorentz
invariant, the nucleons move independently in
mean fields which originate from the nucleon-nucleon interaction.
Conditions of causality and Lorentz invariance impose that the
interaction is mediated by the
exchange of point-like effective mesons, which couple to the nucleons
at local vertices. The single-nucleon dynamics is described by the
Dirac equation
\begin{equation}
\label{statDirac}
\left\{-i\mbox{\boldmath $\alpha$}
\cdot\mbox{\boldmath $\nabla$}
+\beta(m+g_\sigma \sigma)
+g_\omega \omega^0+g_\rho\tau_3\rho^0_3
+e\frac{(1-\tau_3)}{2} A^0\right\}\psi_i=
\varepsilon_i\psi_i.
\end{equation}
$\sigma$, $\omega$, and
$\rho$ are the meson fields, and $A$ denotes the electromagnetic potential.
$g_\sigma$ $g_\omega$, and $g_\rho$ are the corresponding coupling
constants for the mesons to the nucleon.
The lowest order of the quantum field theory is the {\it
mean-field} approximation: the meson field operators are
replaced by their expectation values. The sources
of the meson fields are defined by the nucleon densities
and currents.  The ground state of a nucleus is described
by the stationary self-consistent solution of the coupled
system of Dirac and Klein-Gordon equations.

In addition to the self-consistent mean-field
potential, pairing correlations have to be included in order to
describe ground-state properties of open-shell nuclei.
In the framework of the
relativistic Hartree-Bogoliubov model,
the ground state of a nucleus $\vert \Phi >$ is represented
by the product of independent single-quasiparticle states.
These states are eigenvectors of the
generalized single-nucleon Hamiltonian which
contains two average potentials: the self-consistent mean-field
$\hat\Gamma$ which encloses all the long range particle-hole ({\it ph})
correlations, and a pairing field $\hat\Delta$ which sums
up the particle-particle ({\it pp}) correlations.
In the Hartree approximation for
the self-consistent mean field, the relativistic
Hartree-Bogoliubov equations read
\begin{eqnarray}
\label{equ.2.2}
\left( \matrix{ \hat h_D -m- \lambda & \hat\Delta \cr
		-\hat\Delta^* & -\hat h_D + m +\lambda} \right)
\left( \matrix{ U_k({\bf r}) \cr V_k({\bf r}) } \right) =
E_k\left( \matrix{ U_k({\bf r}) \cr V_k({\bf r}) } \right).
\end{eqnarray}
where $\hat h_D$ is the single-nucleon Dirac
Hamiltonian (\ref{statDirac}), and $m$ is the nucleon mass.
The chemical potential $\lambda$  has to be determined by
the particle number subsidiary condition in order that the
expectation value of the particle number operator
in the ground state equals the number of nucleons.
$\hat\Delta $ is the pairing field. The column
vectors denote the quasi-particle spinors and $E_k$
are the quasi-particle energies.
The RHB equations are solved self-consistently, with
potentials determined in the mean-field approximation from
solutions of Klein-Gordon equations
for the sigma meson, omega meson, rho meson and photon
field, respectively.

%
\section{Ground-state properties of A=20 isobars}

The details of the ground-state properties of the $A=20$ isobaric
sequence will depend on the choice of the effective nuclear
interaction. In the last few years several new and more accurate 
parameter sets of meson masses and meson-nucleon coupling constants
for the effective mean-field Lagrangian have been derived. 
Due to large uncertainties in the experimental data, 
older effective interactions were not specifically designed
to describe ground-state isovector properties. Only more
recently have isovector properties been included in the set
of data on which the effective interactions are adjusted.
In the relativistic mean-field model, perhaps the most
accurate set of meson masses and meson nucleon coupling
constants is NL3~\cite{LKR.97}. In most recent applications of
the RHB theory we have used  the NL3 effective interaction
for the mean-field Lagrangian, and the pairing field has been
described by the pairing part of the Gogny interaction
with the D1S parameter set \cite{BGG.84}.
This force has been very carefully adjusted to the pairing
properties of finite nuclei all over the periodic table.
In particular, the basic advantage of the Gogny force
is the finite range, which automatically guarantees a proper
cut-off in momentum space. 

The RHB equations are solved self-consistently, with
potentials determined in the mean-field approximation from
solutions of Klein-Gordon equations for the meson fields.
The Dirac-Hartree-Bogoliubov equations and the equations for the
meson fields are solved by expanding the nucleon spinors
$U_k({\bf r})$ and $V_k({\bf r})$,
and the meson fields in terms of the eigenfunctions of a
deformed axially symmetric oscillator potential.
The calculations for the present analysis have been performed
by an expansion in 12 oscillator shells for the fermion fields,
and 20 shells for the boson fields.
A simple blocking procedure is used in the calculation of
odd-proton and/or odd-neutron systems. The blocking calculations
are performed without breaking the time-reversal symmetry.
A detailed description of the Relativistic Hartree-Bogoliubov
model for deformed nuclei can be found in Ref. \cite{LVR.99}.

In Fig.~\ref{figA} we compare the RHB NL3+D1S binding energies 
of the $A = 20$ nuclei
($^{20}$N, $^{20}$O, $^{20}$F, $^{20}$Ne, $^{20}$Na, $^{20}$Mg)
with experimental data~\cite{AW.95}. The agreement between theory
and experiment is very good. The self-consistent RHB neutron 
and proton ground-state density distributions for the 
$A = 20$ isobaric sequence are shown in Fig.~\ref{figB}. 
We will eventually compare the calculated radii with 
experimental values, but here we notice the trend: a 
pronounced proton skin is observed in $^{20}$Mg, it slowly
disappears and the proton and neutron density distributions 
are almost identical in $^{20}$Ne, the neutron skin develops
and is very pronounced in $^{20}$N. This trend is in 
complete agreement with experimental evidence for the 
existence of a proton skin for $^{20}$Mg and of a neutron 
skin for $^{20}$N~\cite{Chul.96}.

The calculated ground-state quadrupole deformations of
$A = 20$ isobars are shown in Fig.~\ref{figC} 
as function of the isospin projection $T_z$.
Except for $^{20}$Ne and $^{20}$Na, the $A = 20$ nuclei
are essentially spherical. $^{20}$Ne is strongly prolate deformed
($\beta_2 = 0.365$), and an oblate ground-state is calculated 
for $^{20}$Na ($\beta_2 = -0.153$). 
The surface thickness and diffuseness parameters of the 
proton and neutron density distributions (Fig.~\ref{figA}) are
displayed in Fig.~\ref{figD}. The
surface thickness $s$ is defined to be the change in radius
required to reduce $\rho (r) / \rho_0$ from 0.9 to 0.1
($\rho_0$ is the density in the center of the nucleus). The
diffuseness parameter $\alpha$ is determined by fitting the
neutron density profiles to the Fermi distribution
\begin{equation}
\rho (r) =  {\rho_0} \left (1 + exp({{r - R_0}\over
\alpha})\right)^{-1} ,
\end{equation}
where $R_0$ is the half-density radius.
The surface thickness and diffuseness parameters quantify 
the proton and neutron skins: for $^{20}$Mg $s(p) = 1.81$ fm,
$s(n) = 1.50$ fm, and the corresponding diffuseness 
parameters $\alpha(p) = 0.42$ fm, $\alpha(n) = 0.35$ fm;
for $^{20}$N $s(p) = 1.60$ fm, $s(n) = 1.94$ fm, 
$\alpha(p) = 0.38$ fm, $\alpha(n) = 0.46$ fm.
For $^{20}$Ne the surface thickness and diffuseness parameters
of the proton and neutron distributions are identical.

The calculated radii are compared with experimental 
data~\cite{Chul.96, Suz.97} in Figs.~\ref{figE} and \ref{figF}.
In Fig.~\ref{figE} we display 
the nuclear matter radii (left panel), and the differences between 
proton and neutron radii (right panel) of the $A = 20$ nuclei as 
functions of the isospin projection $T_z$. The theoretical values
are in very good agreement with experimental data. The calculated
matter radii reproduce the observed staggering between even and odd 
values of $T_z$. The only serious discrepancy is $^{20}$O: 
the calculated matter radius is 2.79 fm and the experimental value
is 2.69(3)~\cite{Chul.96}. However, as we have already mentioned 
in the introduction, several theoretical models have failed to
reproduce this extremely small value for the matter radius 
of $^{20}$O~\cite{BH.96,KTS.97,Des.98,SFB.99}, and it was also
suggested that the reported experimental value should be
reconsidered~\cite{SFB.99}. All theoretical calculations predict  
the matter radius in the interval between 2.77 fm and 2.83 fm, 
in agreement with the result of the present analysis. In fact,
since most models predict over-binding for the O isotopes  
($^{26}$O, and even $^{28}$O, are bound nuclei in most model
calculations), one would expect the theoretical value
for the matter radius in $^{20}$O to be smaller than the
experimental one. With the exception of $^{20}$O, also
the calculated differences between proton and neutron radii 
(right panel) are in very good agreement with experimental
data. In particular, the theoretical values agree with the
large differences found in nuclei with a large excess of 
neutrons ($^{20}$N, $0.33\pm 0.15$ fm) or protons
($^{20}$Mg, $0.50\pm 0.28$ fm).

Finally, in Fig.~\ref{figF} theoretical and experimental
values are compared for the proton radii as function 
of the isospin projection $T_z$ (left panel), and 
for the matter radii as function of the binding energy. 
We notice a very good agreement for all the proton radii.
The RHB model also reproduces, with the above discussed 
exception of $^{20}$O, the functional dependence of the
matter radii on the binding energy of the $A = 20$ nuclei.
\bigskip

%
\section{Summary}

In this study we have applied the Relativistic Hartree-Bogoliubov
theory in an analysis of ground-state properties of nuclei that
belong to the $A = 20$ isobaric sequence. The NL3 effective interaction
has been used for the mean-field Lagrangian, and pairing correlations
have been described by the pairing part of the finite range Gogny 
interaction D1S. This particular combination of effective forces 
in the $ph$ and $pp$ channels has been used in most of our 
recent applications of the RHB theory. Theoretical predictions for
binding energies, 
neutron and proton ground-state density distributions,
quadrupole deformations, nuclear matter radii, proton radii
and differences between proton and neutron radii for
$^{20}$N, $^{20}$O, $^{20}$F, $^{20}$Ne, $^{20}$Na, and $^{20}$Mg
have been analyzed and compared with available experimental data.
It has been shown that the NL3 effective force provides a
very good description of the observed ground-state properties as
function of the isospin projection $T_z$. The results of the 
present analysis confirm that the isovector channel
of the NL3 interaction is correctly parameterized
and that this effective force can be used to describe
properties of nuclei far from $\beta$-stability.

\bigskip
\bigskip

\leftline{\bf ACKNOWLEDGMENTS}

This work has been supported in part by the
Bundesministerium f\"ur Bildung und Forschung under
project 06 TM 979, by the Deutsche Forschungsgemeinschaft,
and by the Gesellschaft f\" ur Schwerionenforschung (GSI) Darmstadt.

\newpage

\begin{figure}
\caption{
Binding energies of $A = 20$ nuclei calculated 
with the NL3 + Gogny D1S effective interaction.
The theoretical values are compared with the experimental
binding energies~\cite{AW.95}.}
\label{figA}
\end{figure}

\begin{figure}
\caption{
Self-consistent RHB neutron and proton 
ground-state density distributions of $A = 20$ nuclei.} 
\label{figB}
\end{figure}

\begin{figure}
\caption{
Calculated ground-state quadrupole deformations of 
$A = 20$ isobars as function of the isospin projection $T_z$.}
\label{figC}
\end{figure}

\begin{figure}
\caption{
Surface thickness of the neutron and proton density
distributions of the $A = 20$ isobars, calculated 
with the NL3 + Gogny D1S effective interaction.
In the insert the corresponding
values for diffuseness parameter are included.}
\label{figD}
\end{figure}

\begin{figure}
\caption{
The nuclear matter radii (left), and the differences between 
proton and neutron radii (right) of the $A = 20$ nuclei as 
functions of the isospin projection $T_z$. Results of fully
self-consistent RHB calculations are compared with experimental
data~\cite{Chul.96}.}
\label{figE}
\end{figure}
 
\begin{figure}
\caption{
Theoretical values for the proton radii as function of the isospin 
projection $T_z$ (left), and for the matter radii as function of the
binding energy (right) of the $A = 20$ isobaric sequence, in 
comparison with experimental data~\cite{Chul.96}.}
\label{figF}
\end{figure}

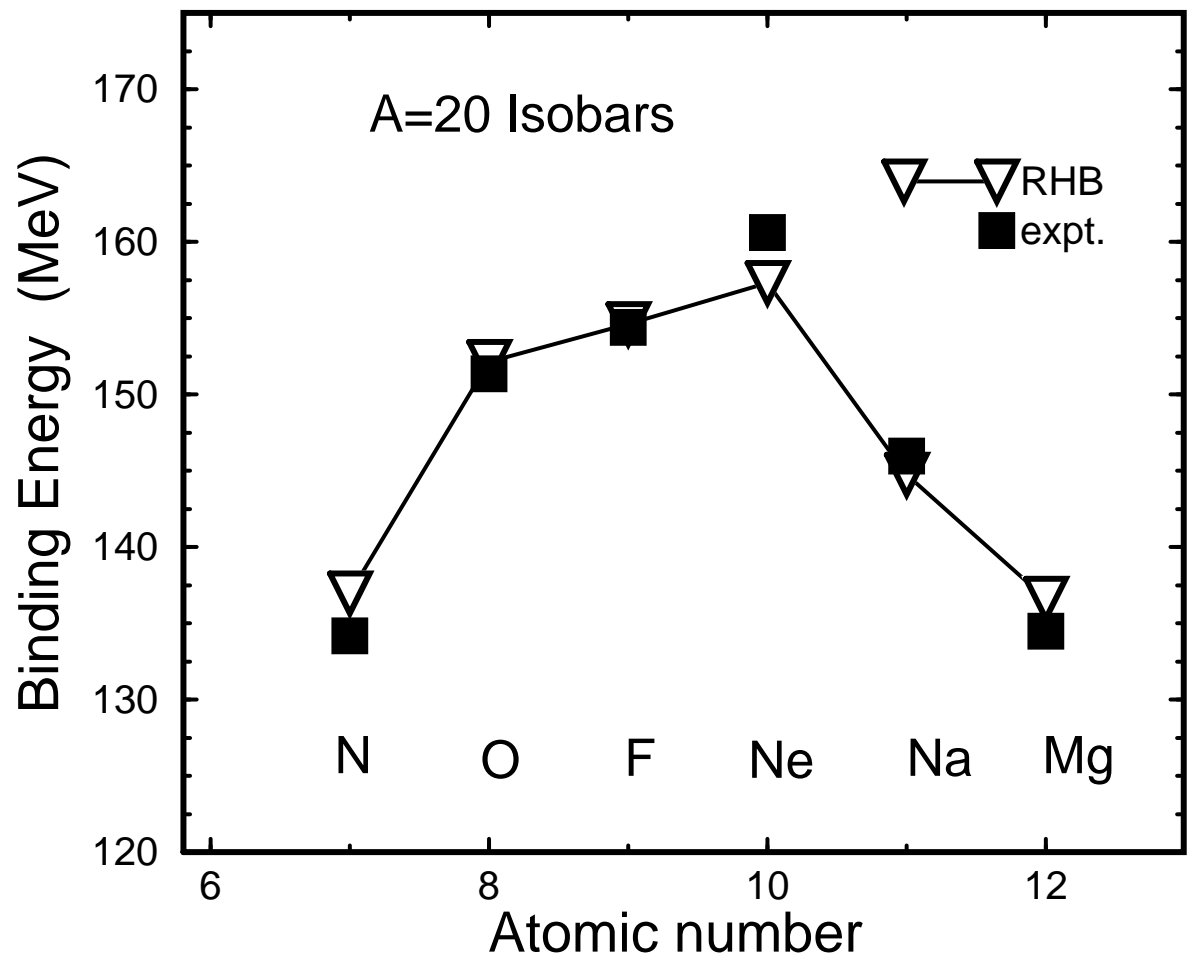
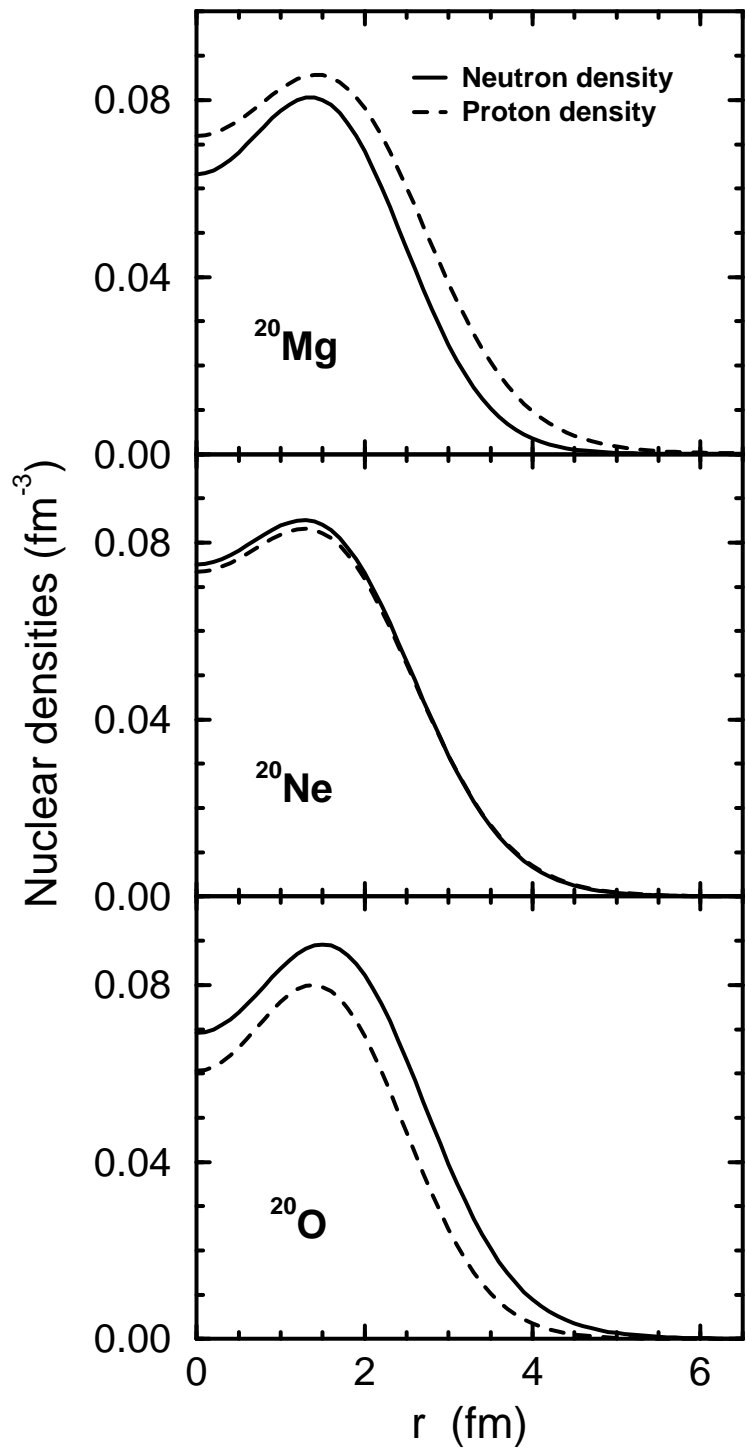
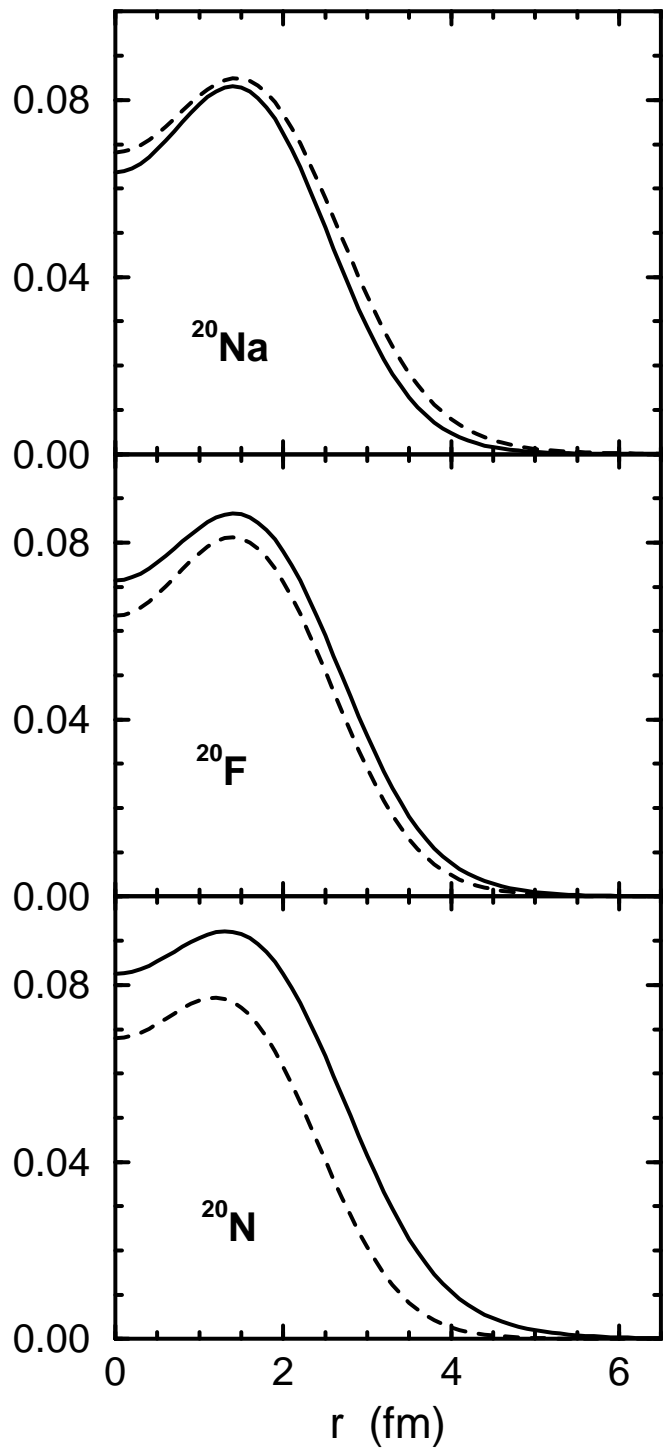
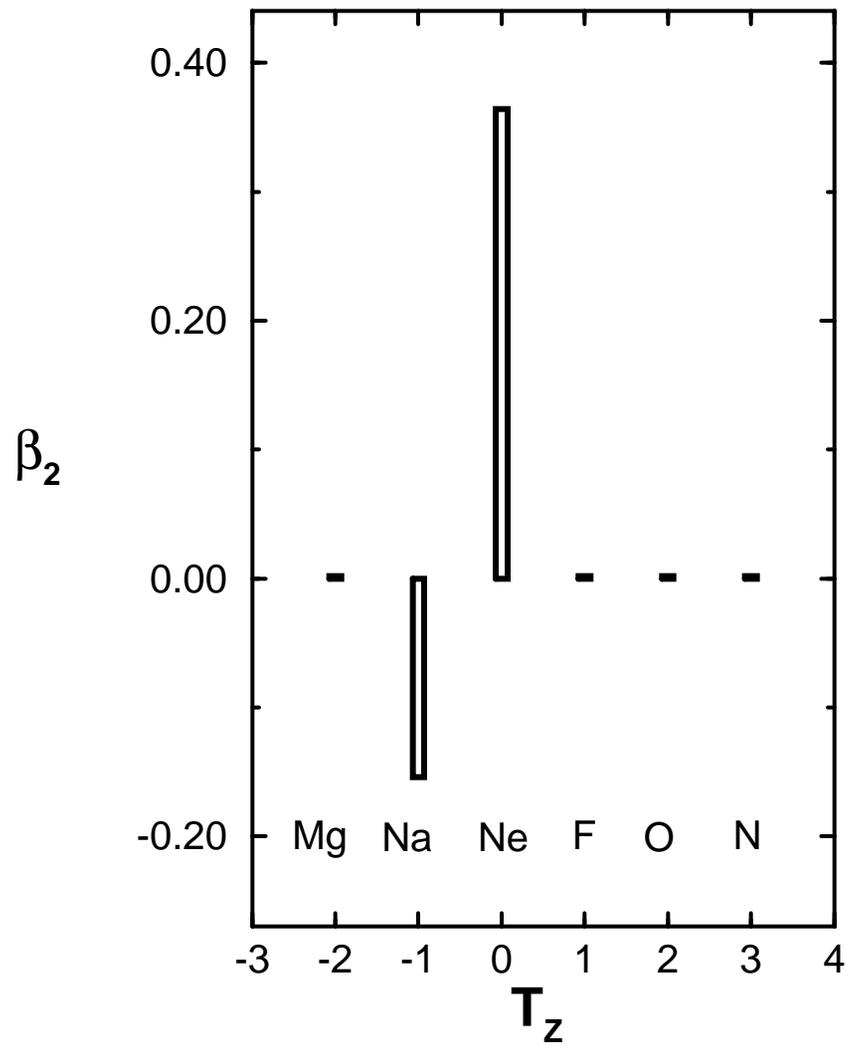
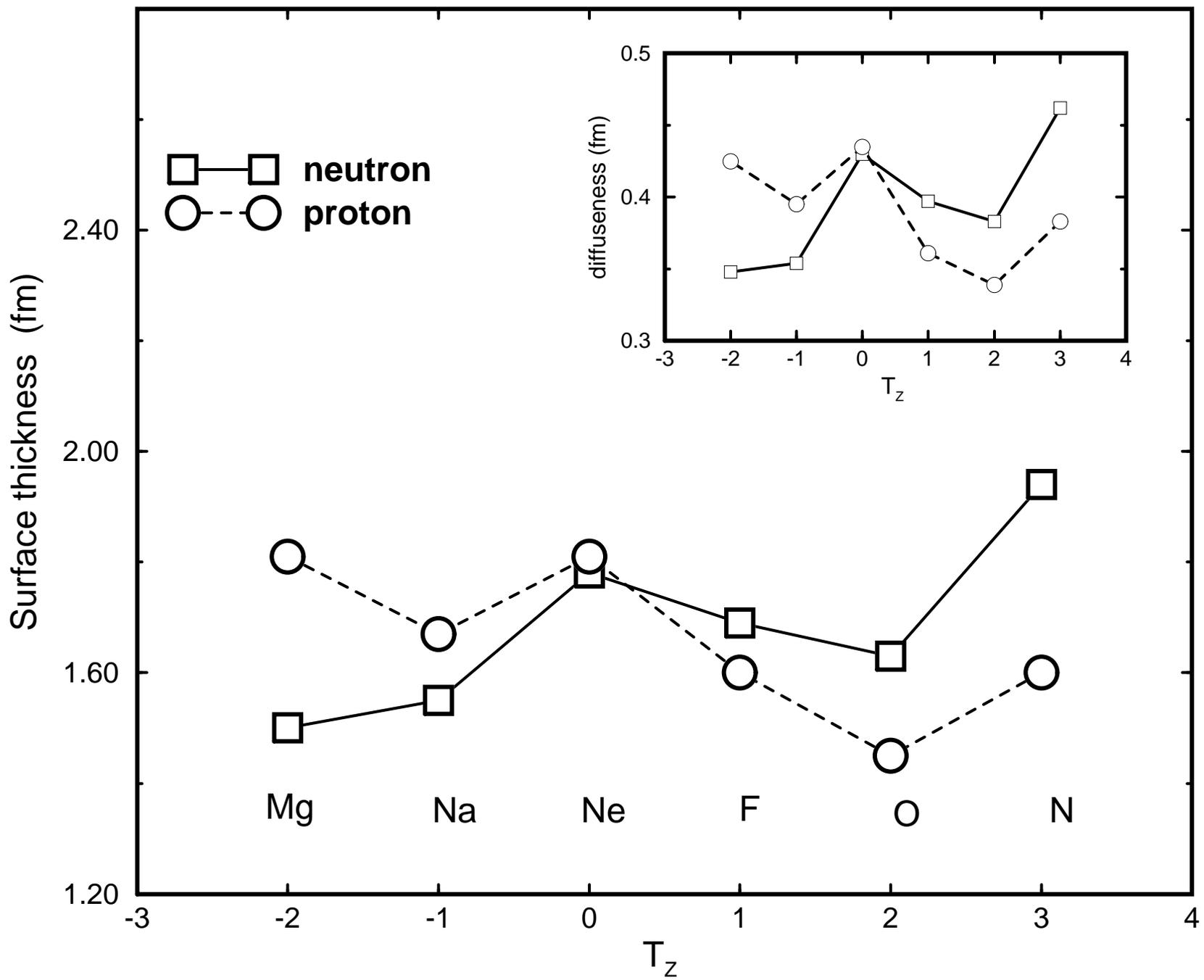
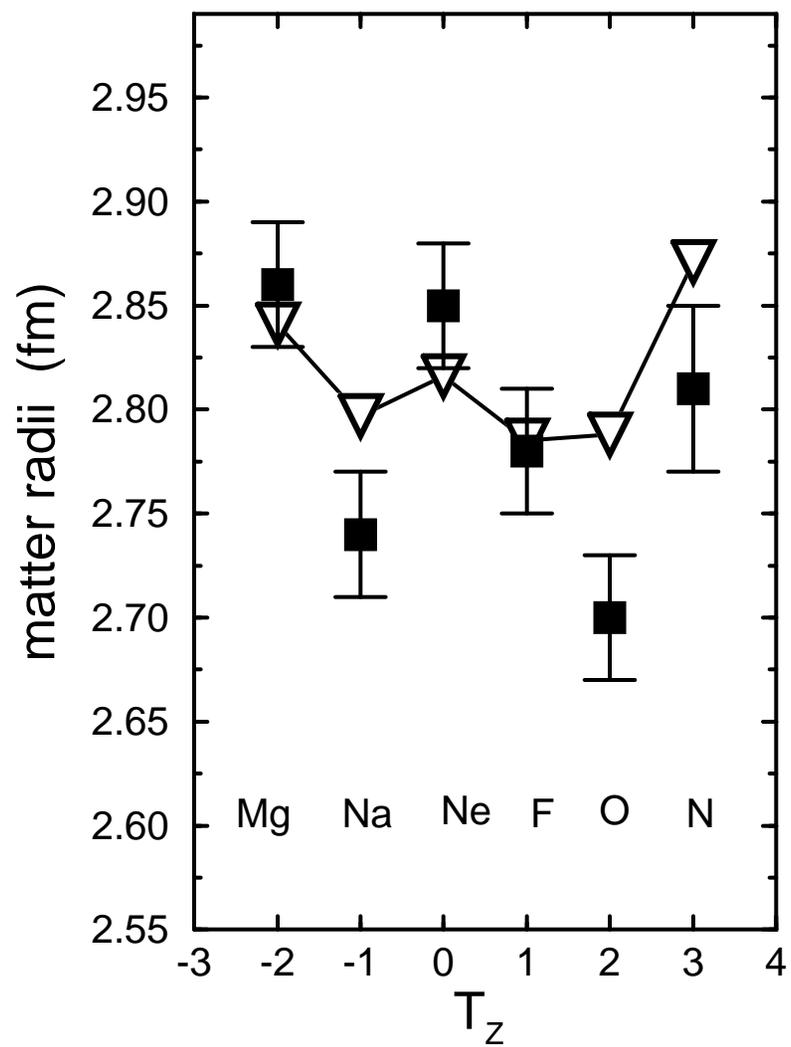
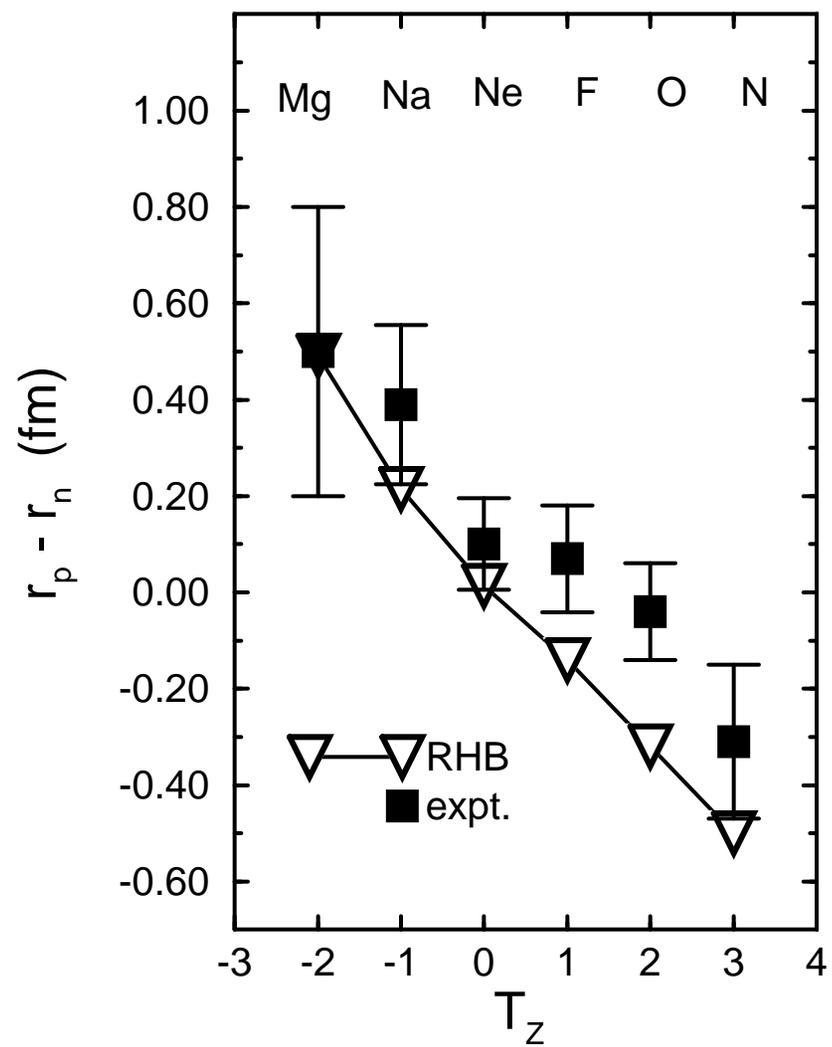
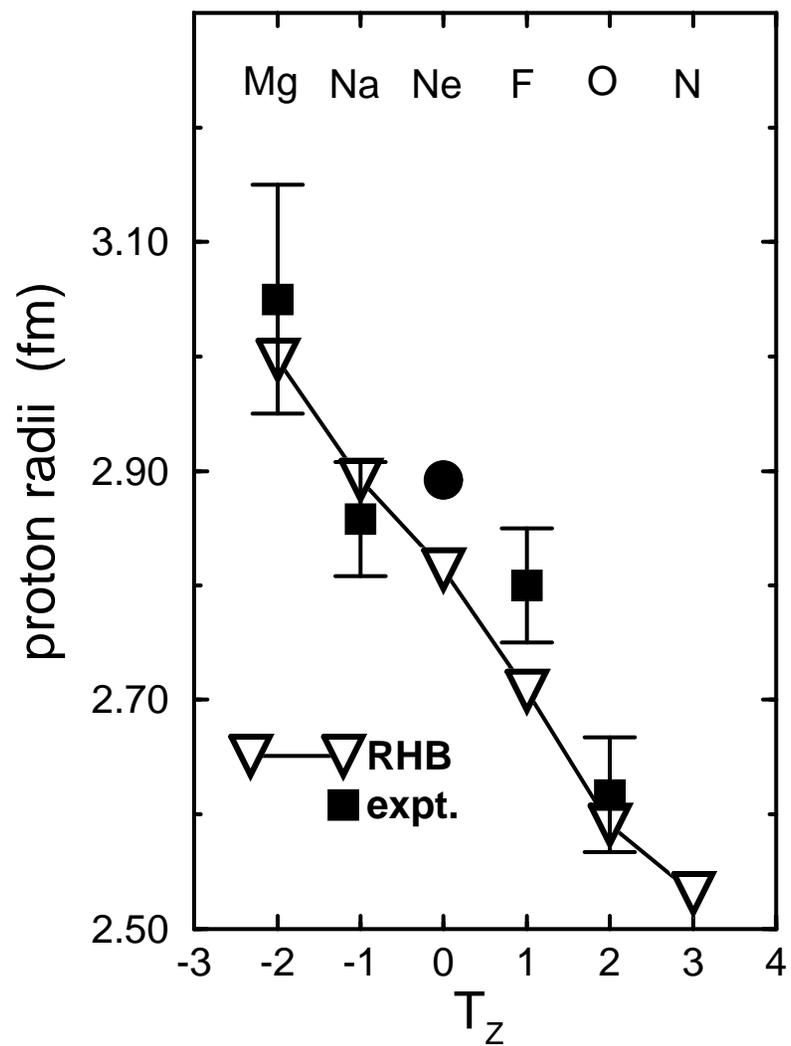
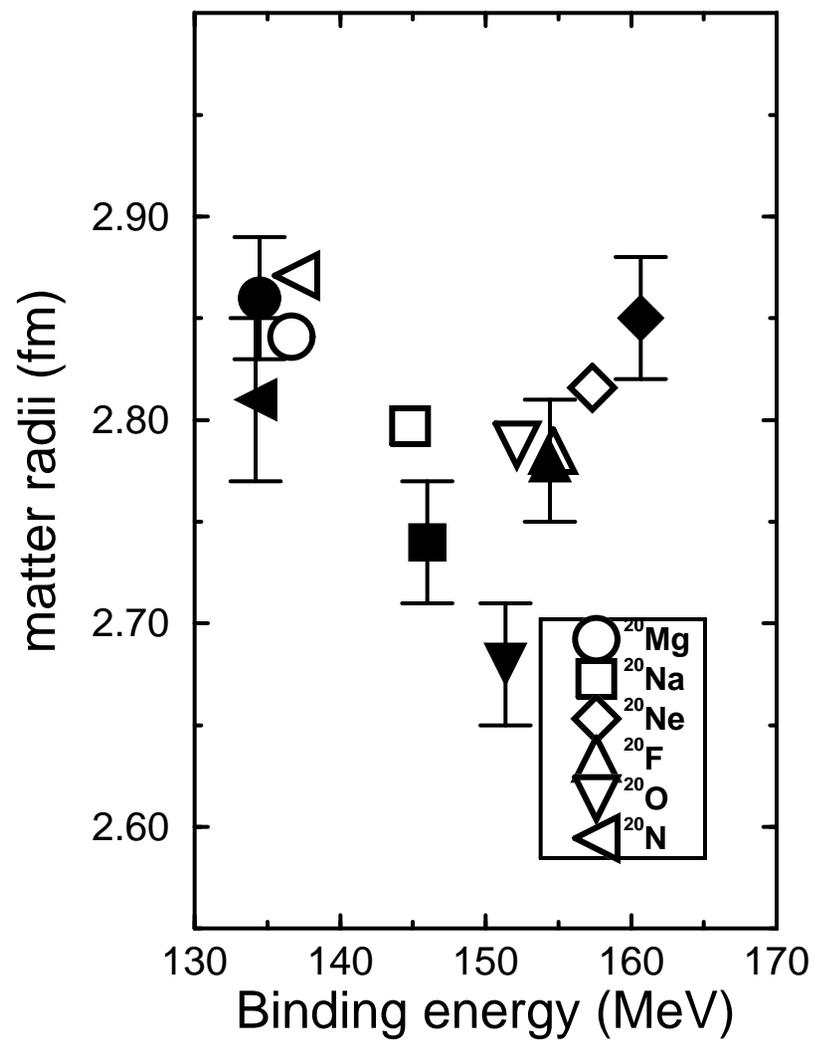
\end{document}